\documentclass[a4paper,11pt]{article}

\usepackage[normalem]{ulem}
\usepackage{jcappub} 
\usepackage{amsmath}
\usepackage{amssymb}
\usepackage{color}
\usepackage{graphicx}
\usepackage{comment}
\usepackage{multirow}
\usepackage[dvipsnames]{xcolor}

\usepackage[utf8]{inputenc}

\newcommand{\lp}{\left(}
\newcommand{\rp}{\right)}
\newcommand{\lb}{\left[}
\newcommand{\rb}{\right]}

\usepackage{natbib}
\usepackage{amsmath}
\usepackage{amssymb}

\begin{document}

\title{\boldmath Cosmological dynamics of extended chameleons}

\author[a]{Nicola Tamanini}
\emailAdd{nicola.tamanini@cea.fr}
\affiliation[a]{Institut de Physique Th\'eorique, CEA-Saclay,
CNRS UMR 3681, Universit\'e Paris-Saclay, F-91191 Gif-sur-Yvette, France}

\author[b]{Matthew Wright}
\emailAdd{matthew.wright.13@ucl.ac.uk}
\affiliation[b]{Department of Mathematics, University College London, Gower Street, London, WC1E 6BT, UK}

\abstract{
	We investigate the cosmological dynamics of the recently proposed extended chameleon models at both background and linear perturbation levels.
  Dynamical systems techniques are employed to fully characterize the evolution of the universe at the largest distances, while structure formation is analysed at sub-horizon scales within the quasi-static approximation.
  The late time dynamical transition from dark matter to dark energy domination can be well described by almost all extended chameleon models considered, with no deviations from $\Lambda$CDM results at both background and perturbation levels.
  The results obtained in this work confirm the cosmological viability of extended chameleons as alternative dark energy models.
}

\maketitle


\section{Introduction} 
\label{sec:introduction}

Scalar field theories are among the most promising alternative explanations of the late time acceleration of the universe, and in fact they account for several popular models of dark energy (see e.g.~\cite{Copeland:2006wr} for a review).
Although the more familiar cosmological constant scenario is well in agreement with all current observations \cite{Betoule:2014frx,Ade:2015xua}, it suffers from different theoretical problems \cite{Weinberg:1988cp,Martin:2012bt,Zlatev:1998tr}, which can be (partially) solved, or at least avoided, introducing new scalar degrees of freedom.
Furthermore from a dynamical perspective, cosmological models with scalar fields offer a wider set of solutions not only for the background evolution of the universe, but also for the growth of structures at the perturbation level.
Not to mention that they are also easier to handle than other dark energy models where higher spin fields appear.

The hypothetical presence of a scalar field at cosmological scales would however give rise to a fifth force between known particles of the Standard Models even at small distances \cite{Carroll:1998zi}.
These new interactions would lead to deviations from the geodesic motion of test masses, which in principle can be detected either by laboratory experiments or by Solar System observations \cite{Joyce:2014kja}.
In order to cope with this problem, scalar field theories where this fifth force is weak, and thus undetectable at small scales, but becomes relevant at large, i.e.~cosmological, distances have been proposed.
The prototype class of these models is represented by the so-called {\it chameleon theories} \cite{Khoury:2003aq,Khoury:2003rn}, whose scalar field is influenced by an effective potential depending on the surrounding matter.
When the environmental matter density is high, such as on the Earth or within the Solar System, the form of the effective potential provides a short range fifth force, while at cosmological scales, where the matter energy density is extremely low, the scalar field mediates a long range force, capable of driving the acceleration of the universe.
Chameleon theories are however not the only ones able to efficiently hiding the scalar field at small scales, but other models, relying on different screening mechanisms, can do the trick: for example the Vainshtein \cite{Vainshtein:1972sx}, symmetron \cite{Hinterbichler:2010es,Hinterbichler:2011ca} and $K$-mouflage \cite{Babichev:2009ee,Brax:2014wla} scenarios.

In a recent work \cite{Brax:2015fcf} a completely new ballpark for screening a scalar field at small scales has been proposed.
This new paradigm naturally generalises the standard chameleon theories, and thus all models encompassed within this class have collectively been named {\it extended chameleons} (ECs).
The characteristic feature of ECs resides in the complete freedom of choosing the dependence on the matter environment of the mass and minimum of the scalar field effective potential.
Thus, unlike in standard chameleon and symmetron theories, where the mass and minimum are related through the full non linear form of the effective potential, with ECs one has the freedom to vary these two quantities with no restriction, enlarging the possible theoretical space in which screening models can be created.
This has been achieved employing the original framework of Scalar-Fluid theories \cite{Boehmer:2015kta,Boehmer:2015sha,Koivisto:2015qua}, where a scalar field is allowed to interact with the matter sector, effectively described as a relativistic fluid, in full generality.
Although Scalar-Fluid theories have originally been used to build new models of interacting dark energy, which actually belong to a wider class of phenomenological models \cite{Tamanini:2015iia}, they can be adapted to any context where a scalar field is coupled to matter: not only screening mechanisms such as the ECs, but also inflationary reheating for example.
The potentialities of Scalar-Fluid theories are far from being completely understood, and further work on the subject will be required.

To come back to ECs, in \cite{Brax:2015fcf} it has been shown that all these models can be restricted by Solar System observations, though, contrary to standard chameleons, a large part of the parameter space remains still viable.
Moreover several EC models present interesting applications at galactic scales, both outside and inside the virial radius, where interesting signatures in the galaxy rotation curves arise and can in principle be compared against observational data.
This in particular happens for a model whose effective mass remains constant as the matter energy density changes.
In this case the only quantity that varies with the environment is the minimum of the effective potential, whose non constancy has been found to be essential for the successful appearance of non negligible effects at small and microscopic scales~\cite{Brax:2015fcf}.

Although the features of EC models have been extensively investigated at both Solar System and galactic scales in \cite{Brax:2015fcf}, the implications at cosmological scales have only been briefly mentioned.
The scope of this paper is thus to study the dynamics of ECs at the background cosmological level, and to analyse the effects on structure formation using linear perturbation theory.
After a brief introduction to Scalar-Fluid theories and ECs (Sec.~\ref{sec:scalar_fluid_theories_and_extended_chameleons}), dynamical systems techniques are employed to characterise the complete background cosmological evolution provided by some specific EC models (Sec.~\ref{sec:dynamical_systems_analysis}).
Then the equations of cosmological perturbations at first order are presented and the implications on sub-horizon structure formation, within the quasi-static approximation, are investigated (Sec.~\ref{sec:perturbations_and_structure_formation}).
A discussion summarising all the main results is provided at the end (Sec.~\ref{sec:conclusions}).

{\it Notation}: In this paper the metric signature is taken to be $(-,+,+,+)$, the speed of light will be set to one $c=1$, and $\kappa^2 = 8\pi G$.


\section{Scalar-Fluid theories and extended chameleons} 
\label{sec:scalar_fluid_theories_and_extended_chameleons}

In this section we will briefly introduce Scalar-Fluid theories, focusing on the details of the theory that will be necessary for this work. We then introduce Extended Chameleon theories within this framework.

\subsection{Scalar-Fluid theories}
Scalar-Fluid theories were first introduced in~\cite{Boehmer:2015kta,Boehmer:2015sha}, as new models of dark energy interacting with the matter sector.
The general action for this class of theories is given by the sum of the following Lagrangian densities
\begin{align}
  \mathcal{S} = \int  d^4x\left(\mathcal{L}_{\rm grav} +\mathcal{L}_{\rm fluid}+ \mathcal{L}_\phi+ \mathcal{L}_{\rm int}\right) \,.
\label{action}
\end{align}
The gravitational sector $\mathcal{L}_{\rm grav}$ is given by the standard Einstein-Hilbert Lagrangian
\begin{align}
  \mathcal{L}_{\rm grav} = \frac{\sqrt{-g}}{2\kappa^2}R \,,
\end{align}
where $R$ is the Ricci scalar of the metric $g_{\mu\nu}$, and $g$ is the metric's determinant. The fluid Lagrangian $\mathcal{L}_{\rm fluid}$ is based on the relativistic fluid Lagrangian description described in~\cite{Brown:1992kc}, and is given by
\begin{align}
  \mathcal{L}_{\rm fluid} = -\sqrt{-g}\,\rho(n,s) + J^\mu\left(\varphi_{,\mu}+s\theta_{,\mu}+\beta_A\alpha^A_{,\mu}\right) \,.
\label{action:fluid}
\end{align}
Here $\rho$ is the energy density of the the fluid, which in general depends on both the particle number density $n$ and the entropy density $s$. $\theta$, $\varphi$ and $\beta_A$ are all Lagrange multipliers of the Lagrangian, where $A$ takes values $1,2,3$, and $\alpha_A$ are the Lagrangian coordinates of the fluid. The vector density particle number $J^\mu$ is related to the particle number density $n$ as 
\begin{align}
  J^\mu=\sqrt{-g}\,n\,u^\mu\,, \qquad |J|=\sqrt{-g_{\mu\nu}J^\mu J^\nu}\,, \qquad n=\frac{|J|}{\sqrt{-g}} \,,
\label{vector density}
\end{align}
where $u^\mu$ is the fluid 4-velocity satisfying the normalisation constraint $u_\mu u^\mu=-1$. 

The scalar field Lagrangian $\mathcal{L}_{\phi}$ is taken to be of the canonical type
\begin{align}
  \mathcal{L}_\phi = -\sqrt{-g}\, \left[\frac{1}{2}\partial_\mu\phi\,\partial^\mu\phi +V(\phi)\right] \,,
\end{align}
where $V$ is a general potential of the scalar field $\phi$. Finally this leaves us to determine the interacting Lagrangian $\mathcal{L}_{\rm int}$. In this work we are interested in algebraic couplings between the fluid and the scalar field, of the type studied in~\cite{Boehmer:2015kta}, as opposed to the derivative couplings considered in~\cite{Boehmer:2015sha}. This means we take the interacting Lagrangian to be of the general form
\begin{align}
  \mathcal{L}_{\rm int} = -\sqrt{-g}\, f(n,s,\phi) \,, \label{027}
\end{align}
with $f$ a general function of its arguments. The specific form of $f$ determines the particular Scalar-Fluid model at hand. In what follows we will assume that $f$ is independent of the entropy density, so that the coupling is adiabatic.

Now let us write down the field equations for this action. Varying~(\ref{action}) with respect to the metric yields the following Einstein field equations
\begin{align}
  G_{\mu\nu} =\kappa^2\left( T_{\mu\nu} +T_{\mu\nu}^{(\phi)} +T_{\mu\nu}^{\rm (int)} \right)\,.
  \label{Einstein}
\end{align}
The three different energy momentum tensors are defined to be
\begin{align}
  T_{\mu\nu} & = p\, g_{\mu\nu} + (\rho+p)\, U_\mu U_\nu \,, \\
  T_{\mu\nu}^{(\phi)} &= \partial_\mu\phi\,\partial_\nu\phi -g_{\mu\nu} \left[\frac{1}{2}\partial_\mu\phi\,\partial^\mu\phi +V(\phi)\right] \,,  \\
  T_{\mu\nu}^{\rm (int)} & = p_{\rm int}\,g_{\mu\nu} + \left(p_{\rm int}+\rho_{\rm int}\right) U_\mu U_\nu \,.
\end{align}
In the above the pressure $p$ of the fluid is defined as
\begin{align}
  p = n\frac{\partial\rho}{\partial n}-\rho \,, 
  \label{eq:002}
\end{align}
whereas the interacting pressure and energy density are defined to be
\begin{align}
  \rho_{\rm int} = f(n,\phi) \quad\mbox{and}\quad p_{\rm int} = n\frac{\partial f(n,\phi)}{\partial n}-f(n,\phi) \,.
\end{align}

For the purposes of this work, we will be interested only in coupling functions $f$ which depend on $n$ only through the energy density. Hence the coupling function takes the particular functional form $f(n,s,\phi)=f(\rho,\phi)$. In such a case the interacting pressure can be written as
\begin{align}
p_{\rm int}=(\rho+p)\frac{\partial f}{\partial \rho}-f \label{pint},
\end{align}
where the relation (obtained from Eq.~\eqref{eq:002})
\begin{align}
\frac{\rho+p}{n}=\frac{\partial\rho}{\partial n} \,,
\end{align}
has been used.

Finally, varying the action with respect to the scalar field gives the Klein-Gordon equation, which reads
\begin{align}
   \Box \phi-\frac{\partial V}{\partial \phi}-\frac{\partial f}{\partial \phi}=0\, , \label{KGeqn}
\end{align}
where $\Box=\nabla^\mu\nabla_\mu$, and $\nabla_\mu$ is the covariant derivative of the metric $g_{\mu\nu}$. This means that the scalar field feels an effective potential $V_{\rm eff}$ of the form
\begin{align}
V_{\rm eff}=V(\phi)+f(\rho,\phi),
\end{align}
which depends on both the scalar field and the local energy density.

\subsection{Extended Chameleons}

Now that we have briefly reviewed the elements of Scalar-Fluid theories, we are ready to introduce the subclass of ECs.
These models are characterised by an effective potential given by
\begin{equation}
  V_{\rm eff}(\rho,\phi) = f(\rho,\phi) = \frac{1}{2} m^2(\rho) \left[ \phi - \phi_0(\rho) \right]^2 \,,
  \label{eq:ext_cham_coupl}
\end{equation}
where $m^2(\rho)$ and $\phi_0(\rho)$ are functions of $\rho$, and the scalar field self-interacting potential has been set to zero.
This represents an effective square potential for the scalar field where both the mass and the vacuum expectation value (the minimum) vary with $\rho$.
It also well approximates any non-linear $V_{\rm eff}$ for values of $\phi$ around the minimum $\phi_0$, as long as suitable functions $m^2(\rho)$ and $\phi_0(\rho)$ are considered.
This is in fact the case of standard chameleons, whose non-linear effective potential $V_{\rm eff}^{\rm cham} = M^{4+\alpha_c}/\phi^\alpha + \rho \exp(\beta_c \phi)$ is well approximated by \eqref{eq:ext_cham_coupl} with power-law functions for $m^2(\rho)$ and $\phi_0(\rho)$ with parameters depending on the constants $M$, $\alpha_c$ and $\beta_c$.
As a matter of fact, the actual EC class has been designed to generalise the standard chameleons results by taking arbitrary power-law functions for $m^2(\rho)$ and $\phi_0(\rho)$; see~\cite{Brax:2015fcf}.

In what follows we will thus reduce our analysis to general polynomial ansatz for $m^2$ and $\phi_0$:
\begin{equation}
  m^2(\rho) = A \rho^\alpha  \quad\mbox{and}\quad \phi_0(\rho) = B \rho^\beta \,,
  \label{eq:003}
\end{equation}
where $A$, $B$ are constants of suitable dimensions and $\alpha$, $\beta$ are dimensionless parameters.
In order to have a positive mass and avoid instabilities, we assume $A>0$ leaving negative effective potentials aside.
The effects on Solar System and galactic distances of ECs, as defined by Eq.~\eqref{eq:003}, have been intensively studied in \cite{Brax:2015fcf}, where it has been shown that interesting signatures can arise in the galaxy rotation curves while passing all local and large scale observations.
Their cosmological dynamics though have only been briefly mentioned in \cite{Brax:2015fcf}, and only for one particular model.
In the next sections we will thus further investigate, employing dynamical systems techniques, the complete dynamics of these models at cosmological distances.

In what follows we will focus on models where $\alpha$ and $\beta$ are given by integer or half-integer numbers.
To shorten the notation we will denote the particular EC model with a given $\alpha$ and $\beta$ by $(\alpha, \beta)$-EC.
For example, the specific EC model with $\alpha=0$ and $\beta=1/2$ will be called $(0,\frac{1}{2})$-EC.


\section{Dynamical systems analysis} 
\label{sec:dynamical_systems_analysis}

\subsection{Background Cosmology}
In this section we will analyse the background cosmological dynamics of these extended chameleon models using dynamical systems techniques. To analyse this background cosmology, we choose to work in a homogeneous and isotropic Friedmann-Robertson-Walker (FRW) line element, as  is required by the cosmological principle. The metric thus takes the following form
\begin{align}
ds^2 = -dt^2 + a(t)^2 \left[\frac{dr^2}{1-K r^2} + r^2 \left(d\theta^2+\sin^2 \theta d\phi^2\right)\right] \,,
\end{align}
where $a(t)$ is the cosmological scale factor and the constant $K=-1,0,1$ depending on whether the universe is spatially open, flat or closed respectively. For the purposes of our cosmological analysis we will make the assumption that the universe is spatially flat, as observations dictate, so we set $K=0$ from now on. We will also impose that all dynamical quantities are homogeneous, that is they depend only on the time coordinate $t$. In particular this means that $\phi$ and $\rho$ will be functions of $t$ only. Working in comoving coordinates, the perfect fluid 4-velocity reduces to simply $u^\mu=(-1,0,0,0)$.

Inserting the FRW metric ansatz into the Einstein field equations~(\ref{Einstein}) and the Klein Gordon equation~(\ref{KGeqn}) yields three independent evolution equations: the two Friedmann equations given by
\begin{align}
  3H^2 &= \kappa^2\left(\rho +\frac{1}{2}\dot\phi^2 +V +f(\rho,\phi) \right) \,,
  \label{eqn:frw1}\\
  2\dot H+3H^2 &=-\kappa^2\left(p+\frac{1}{2}\dot\phi^2 -V +p_{\rm int}\right) \,,
  \label{eqn:frw2}
\end{align}
where $H = \dot{a}/{a}$ is the Hubble rate, and the scalar field evolution equation
\begin{align}
\ddot\phi +3H\dot\phi +\frac{\partial V}{\partial\phi} +\frac{\partial f}{\partial\phi} = 0 \,.
\end{align}
In the particular models studied in this paper, with $f(\rho,\phi)$ given by the functional form~(\ref{eq:ext_cham_coupl}),  the interacting pressure $p_{\rm int}$~(\ref{pint}) is given by
\begin{align}
p_{\rm int}=\frac{1}{2} A \rho^{\alpha -1} \left(B \rho^{\beta }-\phi \right) \left[ B \rho^{\beta } ((\alpha +2 \beta ) (p+\rho)-\rho)+\phi  (\rho-\alpha  (p+\rho))\right] . 
\end{align}

As derived in~\cite{Boehmer:2015kta}, the equation of motion for matter is not modified by the presence of the interacting fluid at large cosmological scales
\begin{align}
\dot\rho +3H\left(\rho+p\right) = 0 \,. \label{matterconservation}
\end{align}
In the remainder of this work we will assume a linear equation of state (EoS), $p=w\rho$ with $w$ a constant, called the matter EoS parameter, which is physically constrained to be between $0$ and $1/3$, with  $w=0$ corresponding to (dark) matter and $w=1/3$ corresponding to radiation. Inserting this into~(\ref{matterconservation}) and solving, it is found
\begin{align}
\rho\propto a^{-3(1+w)} \, ,
\end{align}
and so the matter energy density decays independently as to whether we are working in an interacting universe or a non interacting universe.

Looking at the right hand side of the Friedmann equations, we can define an effective energy density and pressure as follows
\begin{align}
\rho_{\rm eff}&=  \rho +\frac{1}{2}\dot\phi^2 +V +f(\rho,\phi) \,, \\
p_{\rm eff}&= p+\frac{1}{2}\dot\phi^2 -V +p_{\rm int}.
\end{align}
This then naturally leads to defining the effective EoS $w_{\rm eff}$, which is provided by the following ratio
\begin{align}
w_{\rm eff}=\frac{p_{\rm eff}}{\rho_{\rm eff}}=\frac{p+\frac{1}{2}\dot\phi^2-V+p_{\rm int}}{\rho+\frac{1}{2}\dot\phi^2+V+f}. \label{effectiveEOS}
\end{align}
The scale factor will then describe an accelerated expansion if the condition $w_{\rm eff}<-1/3$ holds. 

\subsection{Dynamical systems for extended chameleons}

In this subsection we will introduce the dynamical systems techniques used to analyse the background dynamics of the above cosmological models. We first introduce the concept of a three dimensional dynamical system, which is the main interest of this piece of work, although the definitions here readily generalise to two dimensions or higher dimensions. A three dimensional dynamical system, which is an autonomous system of differential equations, is in general given by the three equations
\begin{align}
  x' = f_1(x,y,z), \quad y' = f_2(x,y,z), \quad z' = f_3(x,y,z)\,,
  \label{eq:dynamsystem}
\end{align}
where the $f_i$ are arbitrary functions of the variables $x,y$ and $z$, yet are independent of the time parameter. The prime denotes differentiation with respect to this time parameter.

A {\it fixed point} or {\it critical point} of such a dynamical system is any solution $(x,y,z)=(x_*,y_*,z_*)$ such that $f_i(x_*,y_*,z_*)=0$ for all $i=1,2,3$. Such a system is at rest as all time derivatives are zero, and the system can remain at this critical point potentially indefinitely. To determine the behaviour of the system near a fixed point, one must assess the {\it stability} of the fixed points. For the purposes of this work, we will use only a linear stability analysis of the critical points. To do this one must analyse the Jacobian matrix of partial derivatives
\begin{align}
\mathcal{J}_{ij}=\frac{\partial f_i(x,y,z)}{\partial x_j}, \quad i,j=1,2,3, \quad \textrm{ where} \quad x_j=(x,y,z)
\end{align}
evaluated at each of the critical points. If all of the eigenvalues of this Jacobian matrix are non-zero, then the point is said to be {\it hyperbolic} and one can perform a linear stability analysis. If the point is non-hyperbolic linear stability analysis breaks down and one must perform a more detailed analysis, see for example~\cite{Boehmer:2014vea,TamaniniPhDthesis,Boehmer:2011tp}.
For linear stability analysis, the stability of each of these points is determined by the signs of the eigenvalues of this matrix. The point is stable if all three of the Jacobian matrix's  eigenvalues have a negative real part. If all three of the eigenvalues have positive real part the point is unstable, whereas if there are both positive and negative eigenvalues the point is said to be a saddle point. 

To simplify the analysis, we will begin by working with a vanishing scalar field potential, so only the effective potential for the scalar field remains (a constant scalar field potential, i.e.~a cosmological constant, will be considered in Sec.~\ref{sub:DS_with_CC}). Assuming this, the flat FRW background cosmology of these models can be analysed by introducing the dimensionless variables
\begin{equation}
	\sigma^2 = \frac{\kappa^2 \rho}{3H^2} \,, \qquad x^2 = \frac{\kappa^2 \dot\phi^2}{6 H^2} \,, \qquad y^2 = \frac{\kappa^2 f}{3 H^2} \,. \label{dynamsys:variables}
\end{equation}
These straightforwardly generalise the standard dimensionless variables used to analyse quintessence models~\cite{Copeland:2006wr,Copeland:1997et}. They reduce the Friedmann constraint~(\ref{eqn:frw1}) to simply
\begin{equation}
	1 = \sigma^2 + x^2 + y^2 \,,
\end{equation}
and so one can choose to dynamically analyse only the variables $x$ and $y$, since one can algebraically relate $\sigma$ to $x$ and $y$. 

Using the background cosmological equations~(\ref{eqn:frw2}) and (\ref{KGeqn}), the dynamical system obtained from these variables is
\begin{multline}
	x' = -\lambda\left(\frac{H}{H_0}\right)^{\alpha-1}y\sigma^{\alpha}-\frac{3}{2}\gamma\left(\frac{H}{H_0}\right)^{\alpha+2\beta-1}x y \sigma^{\alpha+2\beta} \\
	    +\frac{3}{2}x(-1+x^2+(-1+\alpha+ w \alpha)y^2+w\sigma^2)\,,
\end{multline}
\begin{multline}
	y' = \lambda\left(\frac{H}{H_0}\right)^{\alpha-1} x \sigma^{\alpha}-\frac{3}{2}\gamma \left(\frac{H}{H_0}\right)^{\alpha+2\beta-1}\sigma^{\alpha+2\beta}(-1+y^2) \\
	 +\frac{3}{2}y((1-\alpha(1+w))(1-y^2)+x^2+w\sigma^2)\,,
\end{multline}
where a prime denote differentiation with respect to $\eta$, which is introduced such that $d\eta=H dt$. The constants $\gamma$ and $\lambda$ are two dimensionless quantities which have been defined as
\begin{equation}
	\gamma = \sqrt{2A} B \beta (1+w) \left(\frac{\sqrt{3} H_0}{\kappa}\right)^{2 \alpha + \beta -1} \quad\mbox{and}\quad \lambda= \sqrt{A} \left(\frac{\sqrt{3}}{\kappa}\right)^{\alpha} H_0^{ \alpha-1} \,.
\end{equation}

The dynamical system above can be reduced to a two dimensional system only in the case when the explicit dependence on $H$ in the above dynamical system vanishes: this happens if and only if both $\alpha=1$ and $\beta=0$. In all the other cases we need to introduce an extra variable in order to close the system, and so the system will generically be three dimensional. To do this, we introduce the compact variable
\begin{equation}
	z = \frac{H_0}{H_0 + H} \,,
  \label{eq:z_var}
\end{equation}
where $H_0$ is a constant which can be chosen to coincide with the current Hubble rate. This definition ensures that $z$ lies in the range $0\leq z\leq 1$. 

With this additional variable, the dynamical system becomes
\begin{multline}
    x'=-\lambda y \left(\frac{1-z}{z}\right)^{\alpha-1}\sigma^\alpha \\
    +\frac{3}{2}x \left[\gamma y \left(\frac{1-z}{z}\right)^{\alpha+2\beta-1}\sigma^{\alpha+2\beta}+x^2-1+(\alpha(w+1)-1)y^2+w\sigma^2 \right] \label{ds1}
\end{multline}
\begin{multline}
    y'= \lambda\left(\frac{1-z}{z}\right)^{\alpha-1} x \sigma^{\alpha} -\frac{3}{2}\gamma \left(\frac{1-z}{z}\right)^{\alpha+2\beta-1}\sigma^{\alpha+2\beta}(-1+y^2) \\
    	 +\frac{3}{2}y \left[ (1-\alpha(1+w))(1-y^2)+x^2+w\sigma^2 \right] \,, \label{ds2}
\end{multline}
\begin{multline}
	z' =\frac{3}{2}\left[ -\gamma y (1-z)^{\alpha+2\beta}z^{2-\alpha-2\beta}\sigma^{\alpha+2\beta}+(1-z)z\left(1+x^2 +(\alpha(w+1)-1)y^2+w\sigma^2 \right) \right]  \, . \label{ds3}
\end{multline}
We note that this system is invariant under the simultaneous transformation
\begin{align}
y \rightarrow -y, \quad \lambda \rightarrow -\lambda, \quad \gamma \rightarrow -\gamma .
\end{align}
This means we can restrict the analysis to just looking at positive values of one of these variables, since if we want to analyse dynamics for the negative value we can simply reflect the system in a suitable hyperplane. However, looking at the definition of $\lambda$, we see that it is required to be positive, and thus we will only consider positive values of this constant. This, along with the Friedmann constraint implies that the phase space of the system is simply a cylinder with a unit circle base and unit height. 

We find that the effective EoS~(\ref{effectiveEOS}) is given in terms of these new dimensionless variables as
\begin{align}
w_{\rm eff}=x^2+\left( \alpha(w+1)-1\right)y^2+w\sigma^2-\gamma y \sigma^{\alpha+2\beta} \left(\frac{1-z}{z}\right)^{\alpha+2\beta-1}.
\end{align}
We note that depending on the values of the parameters $\alpha$ and $\beta$, the effective EoS is potentially divergent on either the $z=1$ plane or the $z=0$ plane.

Now for generic values of $\alpha$ and $\beta$ the system is practically impossible to analyse as the behaviour of the dynamical system changes extensively depending on their values. And so to proceed we will choose some simple physically interesting values for $\alpha$ and $\beta$ that lead to a dynamical system which is possible to analyse. 


\subsection{Two dimensional model: $(1,0)$-EC}

Let us first analyse the case when the dynamical system reduces to two dimensions, namely the model $(1,0)$-EC. In this case $\alpha=1$ and $\beta=0$, and so the vacuum expectation value of the scalar field is independent of $\rho$ but the mass of the scalar field depends linearly on the energy density.
In \cite{Brax:2015fcf} it has been shown that at small scales only a negligible, for all practical purposes, fifth force is predicted when $\beta = 0$, meaning that all observations and experiments within the Solar System are automatically satisfied for this model.
For $(1,0)$-EC, the Friedmann constraint implies that the phase space of the system is simply the unit circle. In order to simplify the analysis, we will impose that the matter EoS is given by the (dark) matter EoS $w=0$. 

\begin{table}
\begin{center}
\begin{tabular}{|c|c|c|c|}
  \hline
  Point & $x$ & $y$ &  $w_{\rm eff}$  \\
  \hline
  \hline
  $A_{\pm}$ & $\pm 1$ & 0 &  1 \\
  \hline
  $B$ & 0 & $ \pm 1$ &  0 \\
  \hline
  $C_{\pm}$ & $-\frac{3\gamma}{2\lambda}$  & $\frac{\sqrt{-9 \gamma ^2+4 \lambda ^2\pm\sqrt{81 \gamma ^4-36 \gamma ^2 \left(2 \lambda ^2+9\right)+16 \lambda ^4}}}{2 \sqrt{2} \lambda }$  & 0
    \\
  \hline
\end{tabular}
\end{center}
\caption{Critical points of the two dimensional $(1,0)$-EC model along with the value of the effective EoS at each point.}
\label{crit1}
\end{table}

The critical points of the system are displayed in Tab.~\ref{crit1}, along with their existence conditions and the value of the effective EoS at that point. There are potentially six critical points of the system. The four points $A_{\pm}$ and $B_{\pm}$ exist for all values of the parameters $\lambda$ and $\gamma$. On the other hand the points $C_{\pm}$ appear only for certain choices of the parameters $\lambda$ and $\gamma$, however the conditional analytical expressions for their existence are lengthy so we omit them here.

\begin{table}
\begin{center}
\begin{tabular}{|c|c|}
	\hline
	\mbox{Point}  & Stability \\
	\hline \hline
	$A_{-}$ &  Unstable node:  $-3 \sqrt{2} < 3 \gamma - 2 \lambda < 3 \sqrt{2}$ \\
	&  Saddle node: otherwise \\
	\hline
	$A_{+}$ &  Unstable node: $-3 \sqrt{2} < 3 \gamma + 2 \lambda < 3 \sqrt{2}$ \\
	&  Saddle node: otherwise 
	\\ \hline
	$B_{-}$ & Saddle node \\
	\hline
	$B_{+}$ & Saddle node
	\\
	\hline
	$C_{-}$  &  $\gamma\lesssim5\lambda$
	\\
	\hline
	$C_{+}$ &   Saddle node
	\\
	\hline
\end{tabular}
\end{center}
\caption{Stability of the critical points of the two dimensional (1,0)-EC model.} \label{tab01} 
\end{table}

The stability behaviour of the critical points are displayed in Table~\ref{tab01}. The six potential points are:
\begin{itemize}
\item {\it Points $A_{\pm}$}. These two points are dominated by the kinetic energy of the scalar field, with the effective EoS that of a stiff fluid $w_{\rm eff}=1$. No acceleration is present at these point. The points are either unstable or saddle points depending on whether the magnitude of $3\gamma-2\lambda$ is less than $3\sqrt{2}$ for $A_{-}$ or whether $3\gamma+2\lambda$ is less than $3\sqrt{2}$ for $A_{+}$. Such points are unphysical, yet frequently appear as the early time attractors in quintessence type models; cf.~\cite{Copeland:1997et} for example.
\item {\it Points $B_{\pm }$}. These two points exist for all values of $\lambda$ and $\gamma$ and are dominated by the effective potential energy of the scalar field. The solutions have effective equations of state $w_{\rm eff}=0$, mirroring the matter EoS, and thus these solutions represent {\it scaling solutions}.  The eigenvalues of the Jacobian stability matrix indicate that these points are saddle points for all values of the parameters $\lambda$ and $\gamma$. 
\item {\it Points $C_{\pm}$}. These points only exist for certain values of $\lambda$ and $\gamma$, although the analytic expressions for their existence and stability regions are too lengthy to display here. Numerically plotting these regions, we can observe that the point $C_{-}$ exists if approximately the relation $-\gamma\lesssim2\lambda\lesssim\gamma$ holds, and it is stable if the further condition $\gamma\lesssim5\lambda$ approximately holds. On the other hand the point $C_{+}$ is always a saddle. These points also have an effective EoS $w_{\rm eff}=0$ and so these points are also scaling solutions. 
\end{itemize}

\begin{figure}
\centering
\includegraphics[width=0.4\textwidth]{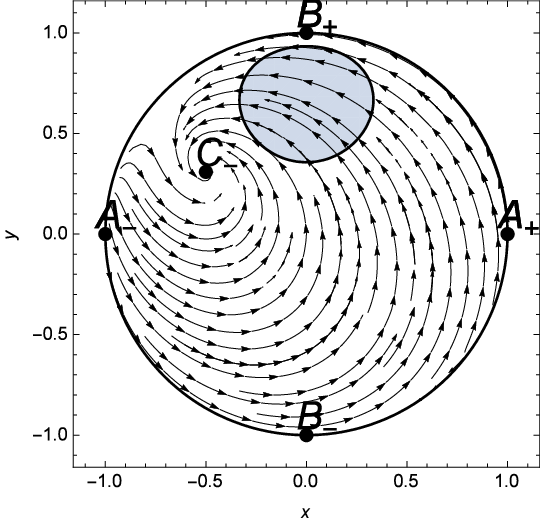}
\hspace{15mm}
\includegraphics[width=0.4\textwidth]{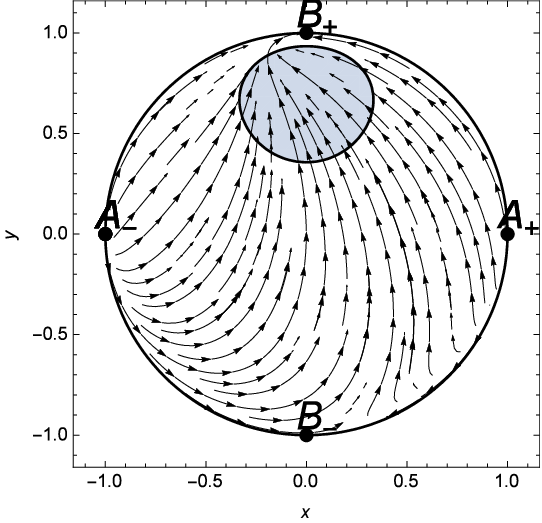}
\caption{Phase space of the two dimensional (1,0)-EC model. The left panel shows trajectories of the dynamical system when the parameters take the values $\lambda=3$, $\gamma=1$. The right hand panel shows the system when $\lambda=1$ and $\gamma=1$. The shaded region indicates the region in which the universe is accelerating, with the effective EoS $w_{\rm eff}<-1/3$, and thus many trajectories exhibit transient periods of acceleration. } \label{2dfigure}
\end{figure}

In Fig.~\ref{2dfigure} example stream plots for two different choices of the parameter values of $\lambda$ and $\gamma$ are displayed. In the left panel the choice $\gamma=1$ and $\lambda=3$ is made.  In this case the critical point $C_-$ exists. Trajectories begin at the unstable stiff matter point $A_-$ before being attracted towards the three saddle points $B_-$, $A_+$ and $B_+$. Many trajectories then pass through a transitory accelerating phase, indicated by the shaded region, before spiralling towards the late time scaling solution attractor $C_-$.  

In the right panel the parameter choice $\lambda=1$ and $\gamma=1$ is made. In this case neither of the critical points $C_\pm$ exist, and we just have the four points $A_{\pm}$ and $B_{\pm}$ lying on the boundary of the phase space. Trajectories begin at either of the unstable nodes $A_+$ or $A_-$. They are then drawn towards the scaling solution at point $B_-$, with then some trajectories undergoing a transient accelerating period, before all trajectories end at the late time attracting solution $B_+$.

An important observation is that there is no critical point which represents a matter dominated epoch, corresponding to $\sigma=1$, located at the origin $(x,y)=(0,0)$ of the two dimensional system. However, all of the points $B_{\pm}$ and $C_{\pm}$ are scaling solutions and mimic the matter dominated EoS.  We note that none of the critical points has an accelerating effective EoS, determined by the condition $w_{\rm eff}<-1/3$, and so a late time acceleration attractor solution is not possible. Thus the cosmology of this model can viably describe our universe if only very fine tuned initial conditions are imposed. In fact, there are regions in phase space in which acceleration can occur, but these periods of accelerations will be transitory.
This means that setting the right initial conditions it might happen that a point in this accelerated region well describes the current observed state of the universe, thought, contrary to what predicted by $\Lambda$CDM, the expansion is doomed to decelerate again in the future.  
Although this might constitute a viable description of our universe, at least at the background level, it also introduces a fine tuning problem much worse that the one arising in $\Lambda$CDM.

\subsection{Three dimensional dynamical systems}
\label{sub:DS_without_CC}

For the remaining values of the parameters $\alpha$ and $\beta$, the system will be three dimensional and  the variable $z$, defined in Eq.~\eqref{eq:z_var}, will be required for the analysis. In this subsection we will analyse mathematically the dynamical system for various different combinations of choices for $\alpha$ and $\beta$ and look for physically interesting cosmology. 

\subsubsection{Positive $\alpha$ and $\beta$}

We will first analyse the dynamics of the system for positive values of the constants $\alpha$ and $\beta$. In the range of values considered, taking into account only integer values $\alpha=0,1$ and half integer values for $\beta=0,1/2,1,2$, we find that the system is not particularly interesting from a cosmological point of view. None of the values considered have stable critical points which have an accelerating effective EoS, implying that only a scenario similar to the one discussed in the two dimensional system case above can be obtained.

\begin{table}
\centerline{
\begin{tabular}{|c|c|c|c|c|c|}
  \hline
  Point & $x$ & $y$ & $z$ & Existence & $w_{\rm eff}$ \\
  \hline
  \hline
  $A_{\pm}$ & $\pm 1$ & 0 & 0 &$ \forall \lambda, \gamma $ & 1 \\
    \hline
  $B_{\pm}$ & 0 & $\pm 1$ & 0 &$ \forall \lambda, \gamma $ & $\alpha-1$ \\
    \hline
  $C_{\pm}$ & $\pm 1$ & 0 & 1 &$ \forall \lambda, \gamma $ & 1 \\
  \hline
  $D_{\pm}$ & 0 & $\pm 1$ & 1 & $ \forall \lambda, \gamma $& $\alpha-1$ \\
  \hline
  $E$ & 0 & 0 & 1 & $ \forall \lambda, \gamma $ &  0 \\
  \hline
  $F$ & 0 & $- \frac{\gamma}{\sqrt{1+\gamma^2}}$ & 0 &  $ \forall \lambda, \gamma $ & $-2\gamma/(2\gamma^2+1)$\\
  \hline
  $\mathcal{C}_1$ & $x$ & $\sqrt{1-x^2}$ & 0 & $\forall \lambda, \gamma$ & Undetermined \\ \hline 
\end{tabular}
\quad
\begin{tabular}{|c|c|c|}
	\hline
	$\alpha$ & $\beta$ & Critical points present \\
	\hline
	\hline
	0 & $\frac{1}{2}$ &  $A_{\pm}$, $B_{\pm}$, $E$, $F$ 
	\\ \hline
	0 & 1 & $E$, $\mathcal{C}_1$ \\
	\hline
	0 & 2 & $E$, $\mathcal{C}_1$ 
	\\
	\hline 
	1 & $\frac{1}{2}$  &  $C_{\pm}$, $D_{\pm}$, $E$, $\mathcal{C}_1$ \\
	\hline
	1 & 1 &  $C_{\pm}$, $D_{\pm}$, $E$, $\mathcal{C}_1$ \\
	\hline
\end{tabular}}
\caption{The left table displays the possible critical points present for the three dimensional systems which have been analysed, along with their existence conditions. The right table displays different choices of the parameters $\alpha$ and $\beta$, and the critical points present in the system for each particular choice.}
\label{3dimensionalcps}
\end{table}

In Table~\ref{3dimensionalcps}, the potential critical points of the three dimensional systems analysed are displayed, along with which values of parameters $\alpha$ and $\beta$ they exist for. The entry $\mathcal{C}_1$ in the table is not a critical point, it is in fact a critical curve (i.e.~every point on the curve is in fact a critical point) which lies on the boundary of the $z=0$ plane. In order to analyse the dynamical system, we will need to be careful with singularities, since the right hand side of the dynamical system~(\ref{ds1})-(\ref{ds3}) is potentially divergent on the $z=1$ plane, depending on the values of $\alpha$ and $\beta$. To regularise the dynamical system, one must multiply the right hand side of the system by an appropriate power of $z$, this simply amounts to a re-parametrisation of the time coordinates.

As none of the models analysed are particularly interesting from a cosmological perspective we will not analyse these models in detail. However some do exhibit some interesting mathematical behaviour and below we briefly discuss the dynamics of each of these systems outlined in Table~\ref{3dimensionalcps}.
\begin{itemize}
\item {\it $\alpha=0$, $\beta=1/2$.} This case corresponds to the physical situation where the mass of the scalar field is constant, but the minimum of the scalar field varies proportional to the square root of the energy density. The model has six critical points, $A_{\pm}$, $B_{\pm}$, $E$ and $F$. This case is possibly the most phenomenologically interesting among the cases analysed here, since one of the critical points, $F$, is a cosmologically accelerating solution for particular values of $\gamma$, with $w_{\rm eff}=-2\gamma^2/(\gamma^2+1)$. However if we look at the eigenvalues of the Jacobian matrix, this point is always found to be a saddle, and thus cannot be the late time attractor solution. Plotting numerically the dynamics on the phase space, it appears as if all trajectories are drawn towards the $z=1$ plane. However, looking at the re-parametrised dynamical system on this plane, it reduces to simply the system
\begin{align}
x'&=-\lambda y \label{circ1}
\\ y'&=\lambda x. \label{circ2}
\end{align}
This is a very simple dynamical system, whose trajectories in phase space are simply circles of constant radius. In this case the critical point $E$ has two purely imaginary eigenvalues, and the point is said to be a centre. This means that once the trajectories reach this plane they start circling, with the effective EoS simply oscillating forever. The points $B_{\pm}$ is also an accelerating de Sitter type solution, however two of the eigenvalues of its Jacobian matrix are always negative, and hence the point is a saddle.

\item{$\alpha=0$, $\beta=1$.} Similar to $(0,\frac{1}{2})$-EC, this model has a constant scalar field mass, but this time the minimum of the scalar field varies linearly with the local energy density.  In this case the system has one critical point $E$ on the origin of the $z=1$ plane, and one critical curve: $\sigma=0$ lying on the boundary of the $z=0$ plane.  Again though, the system always ends up on the $z=1$ plane where the dynamics are given by the restricted system~(\ref{circ1})-(\ref{circ2}) which has circular orbits, with the effective EoS $w_{\rm eff}$ oscillating forever.

\item{$\alpha=0$, $\beta=2$.} In this case the scalar field minimum depends on the square of the energy density. This model is phenomenologically identical to the $\alpha=0$, $\beta=1$ case, with one critical point $E$ and the critical curve $\mathcal{C}_1$. Trajectories are again drawn to the $z=1$ plane with circular orbits. 

\item{$\alpha=1$, $\beta=1/2$.}  This EC model corresponds to the case when the mass of the scalar field varies linearly with the energy density and its minimum varies proportional to the square root of the energy density. This choice of parameters has the five critical points $C_{\pm}$, $D_{\pm}$ and $E$, along with the critical curve $\mathcal{C}_1$. Numerically examining the solutions of the dynamical system, it is seen that the trajectories begin on the critical $z=0$ boundary, the curve $\mathcal{C}_1$. They then are drawn towards the saddle points $C_{\pm}$ and $D_{\pm}$ before arriving at the late time matter dominated attractor $E$. No accelerated critical points are present in this model.

\item{$\alpha=1$, $\beta=1$.} The case $\alpha=\beta=1$ corresponds to the interesting physical situation where both the mass and the vacuum expectation value of the scalar field vary linearly with $\rho$. Similar to the (1,1/2)-EC, the system has the five critical points $C_{\pm}$, $D_{\pm}$ and $E$, and the critical curve $\mathcal{C}_1$. The points $D_{\pm}$  are scaling solutions, that is their effective EoS mirrors that of a matter dominated universe. Examining numerically the behaviour of the trajectories in phase space, it appears that solutions begin on the boundary of the $z=0$ plane, before possibly being drawn towards either of the saddle scaling solutions $D_{\pm}$ and finally arriving at the late time matter dominated solution $E$.

\end{itemize}

\subsubsection{ The $(1,-\frac{1}{2})$-EC model} 

Let us now consider a negative value for the parameter $\beta$, so that the vacuum expectation value of the scalar field depends inversely on the matter density $\rho$. In particular we will look at the $(1,-\frac{1}{2})$-EC, with the scalar mass linearly dependent on the energy density. For this example, the right hand side of the dynamical system~(\ref{ds1})-(\ref{ds3}) diverges on the $z=1$ plane. To remedy this we will multiply the right hand side of the dynamical system by $(1-z)$, this simply amounts to a re-parametrisation of the time coordinate; see e.g.~\cite{TamaniniPhDthesis,Alho:2015ila}.

\begin{table}
\centering
\begin{tabular}{|c|c|c|c|c|c|}
  	\hline
  	Point & $x$ & $y$ & $z$ & Existence & $w_{\rm eff}$  \\
  	\hline
  	$O$ & 0 & 0 & 0 & $\forall \lambda, \gamma$ & 0 \\
	\hline
	$A_{\pm}$ & $\pm1$ & 0 & 0 &  $\forall \lambda, \gamma$ & 1 \\
	\hline
	$B_{\pm}$ & 0 & $\pm 1$ & 0 &  $\forall \lambda, \gamma$ & 0 \\
	\hline
	$C_{\pm}$ & 0 & $\pm 1$ & 1 &  $\forall \lambda, \gamma$ & $\infty$ \\
	\hline
	$G_+$ & 0 & 1 & $\frac{1}{1+ \gamma}$ &  $\gamma>0$  & -1 \\
	\hline
    $G_-$ & 0 & -1 & $\frac{1}{1- \gamma}$ &  $\gamma<0$  & -1 \\
	\hline		
\end{tabular}
\caption{Critical points of three dimensional $(1,-\frac{1}{2})$-EC dynamical system, along with their existence conditions and the value of the effective EoS at that point.}
\label{cp1-12}
\end{table}

The critical points of this model, along with their existence conditions and the value of the effective EoS $w_{\rm eff}$ are displayed in Table~\ref{cp1-12}. This model has nine critical points, with eight existing at any one time. A linear stability analysis of this model is difficult to perform, since the eigenvalues of the Jacobian matrix are infinite at the critical points, except for the origin $O$ which is found to be a saddle. To investigate the behaviour of the dynamical system we will resort to simply looking at numerical plots of the trajectories. 

The system has the following critical points:
\begin{itemize}
\item{\it Point $O$.} The origin of the system is a matter dominated solution, with the effect of the interaction function being zero at this point. This point is always a saddle, and thus leads to a potential matter dominated epoch. 
\item {\it Points $A_{\pm}$}. As in the two dimensional model studied earlier, these two points are unphysical stiff matter solutions with $w_{\rm eff}=1$, which are dominated by the kinetic energy of the scalar field. They lie on the $z=0$ plane and are unstable or saddle points.  
\item {\it Points $B_{\pm }$}.  The points $B_{\pm}$ are scaling solutions in that they have effective EoS $w_{\rm eff}=0$, yet they are not matter dominated, rather their energy is dominated by the potential energy of the interacting function.
\item {\it Points $C_{\pm}$}. These points are dominated by the effective potential energy of the scalar field and have an effective EoS singularity, with $w_{\rm eff}$ diverging. 
\item {\it Points $G_{\pm}$}. These points are also dominated by the effective potential energy of the scalar field. Only one of these two points can exist at any one time, depending on whether $\gamma$ is positive or negative. This point is an accelerating solution, with de Sitter type expansion, $w_{\rm eff}=-1$. It appears from numerical investigations that these points are always stable and are the late time attractor solution. 
\end{itemize}

\begin{figure}
\centering
\includegraphics[width=0.55\textwidth]{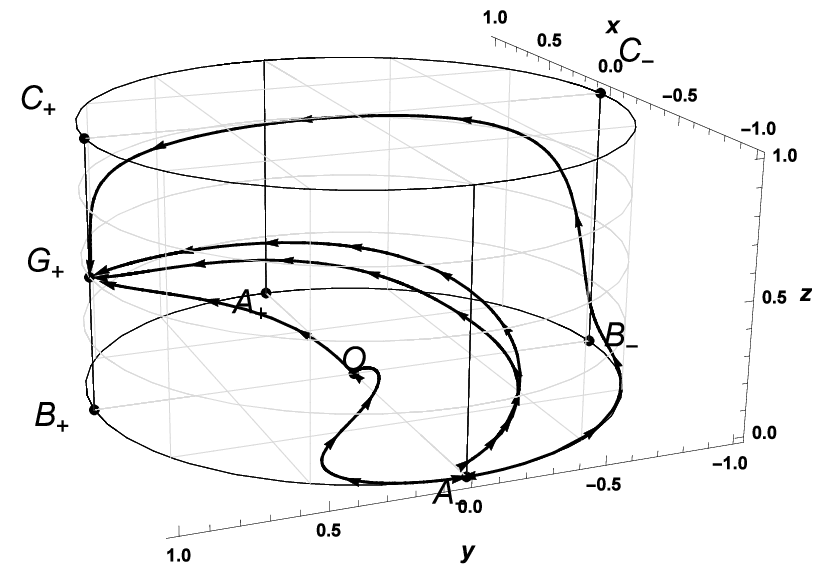}
\caption{Phase space of the three dimensional $(1,-\frac{1}{2})$-EC dynamical system. The parameter choices $\lambda=1$ and $\gamma=1$ were made. The trajectories all evolve to the accelerating attractor solution at $G_{+}$.} \label{3dalpha1beta-12}
\end{figure}

\begin{figure}
\centering
\includegraphics[width=0.55\textwidth]{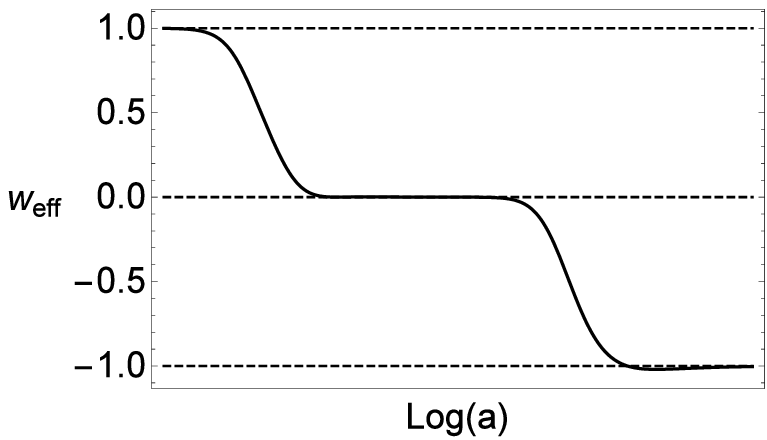}
\caption{Plot of the effective EoS for the particular trajectory which passes through the origin in Figure.~\ref{3dalpha1beta-12}, for the $(1,-\frac{1}{2})$-EC when  $\gamma=1$ and $\lambda=1$. The universe begins in a stiff matter state, before undergoing an era of matter domination. The EoS then transitions to a de Sitter type acceleration, with a brief crossing of the phantom barrier. 
}
\label{weffalpha1beta-12}
\end{figure}

In Figure~\ref{3dalpha1beta-12} the cylindrical phase system along with various trajectories is plotted for the particular choice of parameters $\gamma=1$ and $\lambda=1$. All of the trajectories begin at the stiff matter type unstable point $A_{-}$. Some of them are then drawn towards the various saddle points $B_{-}$, $O$ and $C_+$. They then all arrive at the late time accelerating attractor solution $G_+$. In Figure~\ref{weffalpha1beta-12} the effective EoS for one of these trajectories is plotted over time, in particular the trajectory which passes close to the origin $O$. It is found that at the beginning $w_{\rm eff}$ starts in a stiff state, before passing through an epoch of matter domination. The EoS then settles down at a de Sitter type EoS after briefly crossing the phantom barrier.
A part from the early time stiff fluid dominated phase, this is exactly the cosmological behaviour observed in our universe, with a matter to dark energy transition taking place at late times.
The $(1,-\frac{1}{2})$-EC model thus provides a viable alternative description for dark energy, even without the need of a cosmological constant.

\subsection{Dynamics with cosmological constant}
\label{sub:DS_with_CC}

Many of the models studied so far in this section have either not possessed an accelerating critical point, or these critical points have not been generically stable. In order to look for more potentially viable cosmologies, in this subsection, we will consider the same extended chameleon models as above, but this time we will introduce a cosmological constant into the equations by taking the potential of the scalar field to be given by a simple positive constant, with $V(\phi)=\Lambda$.
Note that a non-vanishing cosmological constant plus a quadratic effective potential of the kind \eqref{eq:ext_cham_coupl}, naturally arise from considering values near the minimum $\phi_0$ of some non linear effective potential, for example $V_{\rm eff} = \Lambda \cosh \left[ m \left( \phi - \phi_0 \right) / \sqrt{\Lambda} \right]$.

Adding $V(\phi)=\Lambda$ into the Friedmann equations modifies the dynamical system. We continue to work with the same variables $\sigma$, $x$ and $y$ as before, however this time it is more appropriate to introduce a redefined $z$, which we take to be
\begin{align}
z^2=\frac{\kappa^2 \Lambda}{3H^2}.
\label{eq:z_var_CC}
\end{align}
This modifies the Friedmann constraint as
\begin{equation}
	1 = \sigma^2 + x^2 + y^2+z^2 \,. \label{friedmannconstraint}
\end{equation}
From the definition of $z$ and the fact that we are working with a positive cosmological constant, we can restrict the analysis to consider only values $z\geq0$. Again working with the variables $x$, $y$ and $z$, with $\sigma$ related to these variables algebraically via~(\ref{friedmannconstraint}), means that the phase space for the dynamical system is simply the upper half unit sphere.  

The dynamical system written in terms of these variables is
\begin{multline}
x'=\frac{1}{2} \Bigg[-2 x \gamma  (w+1) y z^{1-\alpha-2\beta} \sigma^{\alpha +2 \beta } \\ 
+3 x \left(w \sigma^2+y^2 (\alpha +\alpha  w-1)-1\right)- 3 x z^{2} +3 x^3 -2 \lambda  y \sigma^{\alpha } z^{1-\alpha}\Bigg]
\end{multline}
\begin{multline}
y'=\frac{1}{2} \Bigg[-3 y  \left(\alpha -w \sigma^2-x^2+z^2+\alpha  w-1\right) \\ 
+2 z^{1-\alpha-2\beta}\sigma^{\alpha } \left(\gamma  (w+1) \left(y^2+1\right) \sigma^{2 \beta }+\lambda  x z^{2 \beta }\right)+3 y^3 (\alpha +\alpha  w-1) \Bigg]
\end{multline}
\begin{multline}
z'=\frac{1}{2}\left[3 z \left(w \sigma^2+y^2 (\alpha +\alpha  w-1)+x^2+1\right)-2 \gamma  (w+1) y z^{2-\alpha-2\beta}\sigma^{\alpha +2 \beta }-3 z^{3}\right],
\end{multline}
where this time we have introduced the constants $\lambda$ and $\gamma$ as
\begin{align}
\lambda=\frac{\sqrt{3} \sqrt{A} \Lambda ^{\frac{\alpha }{2}-\frac{1}{2}}}{\kappa }, \quad \gamma=\frac{3 \sqrt{A} \beta  B \Lambda ^{\frac{\alpha }{2}+\beta -\frac{1}{2}}}{\sqrt{2}}.
\label{eq:params_cc}
\end{align}

In general the dynamical system will be singular on the $z=1$ plane if either of the conditions $\alpha>1$ or $\alpha+2\beta>1$ hold, in which case we will need to define a new time coordinate and multiply the right hand side of the dynamical system by an appropriate power of $z$ to describe the dynamics on this plane. The effective EoS~(\ref{effectiveEOS}) is given in terms of these variables as
\begin{align}
w_{\rm eff}=x^2+y^2 (\alpha +\alpha  w-1)- z^2+w \sigma^2 -\frac{2}{3} \gamma  (w+1) y z^{1-\alpha-2\beta} \sigma^{\alpha +2 \beta }.
\label{eq:effEoS_cc}
\end{align}
Again, for general $\alpha$ and $\beta$, the system is practically impossible to analyse, however we can make a couple of comments. If $\alpha+2\beta>1$ then generically the effective EoS diverges on the $z=0$ plane. We can also observe that the matter dominated point $(0,0,0)$ is only a critical point of the dynamical system when the condition  $\alpha+2\beta<1$ holds. In what follows we will restrict ourselves to particular physically interesting values of $\alpha$ and $\beta$ for a full analysis of the dynamical system.

\subsubsection{$\alpha=0$ and $\beta=1$}

We will first analyse the dynamical system with the choice of parameters $\alpha = 0$ and $\beta = 1$, namely $(0,1)$-EC. This model represents a new type of screening where the mass of the scalar field remains constant, while the minimum of the effective potential, namely $\phi_0$, varies linearly with the energy density. The physical implications of this model at both Solar System and galactic scales were investigated in~\cite{Brax:2015fcf}. It was found that interesting effects in the galaxy rotation curves can arise while satisfying all constraints on smaller scales.

\begin{table}
\centering
\begin{tabular}{|c|c|c|c|c|}
  \hline
  Point & $x$ & $y$ & $z$ & $w_{\rm eff}$ \\
  \hline
  \hline
  $B$ & 0 & 0 & $1$ & -1 \\
   \hline
   $\mathcal{C}_1$ & $x$ & $\sqrt{1-x^2}$ & 0 & Undetermined
   \\
   \hline
\end{tabular}
\caption{Critical points of the $(0,1)$-EC model with a cosmological constant, along with the values of the effective EoS at each point. }
\label{cpL01}
\end{table}

The dynamical system from a mathematical point of view is particularly simple to analyse. The critical points are displayed in Table~\ref{cpL01}. There is just one critical point $B$ and a critical curve $\mathcal{C}_1$ which is the boundary of the $z=0$ plane. The critical point $B$ located at $(0,0,1)$. This point represents an effective de Sitter type expansion, and is dominated by the energy of the cosmological constant. As expected from such a de Sitter expansion, this point has effective EoS $w_{\rm eff}=-1$. Moreover this point is always the late time attractor: its Jacobian matrix has three negative eigenvalues independently of the values of either $\lambda$ or $\gamma$. 

The boundary, i.e.~the unit circle, of the $z=0$ plane is wholly critical, meaning that every point that lies on it is automatically a critical point.
The past attractor of the phase space is characterised by this unit circle, where the scalar field dominates. Note however that the effective EoS \eqref{eq:effEoS_cc} is not constant along the points of this boundary, but in fact it is undetermined since both $z$ and $\sigma$ go to zero.
Again this reflects the fact that for $(0,1)$-ECs the variables \eqref{dynamsys:variables} and \eqref{eq:z_var_CC} might not constitute the best choice to analyse their cosmological dynamics.
Nevertheless if one assumes that $\sigma^2\gg z$ as trajectories past-approach the curve $\mathcal{C}_1$, which can always be satisfied setting the appropriate initial conditions, then the effective EoS \eqref{eq:effEoS_cc} on the $z=0$ unit circle can approximately be taken as
\begin{equation}
  \left. w_{\rm eff} \right|_{\sigma=0\ \&\ z=0} \approx x^2 - y^2 \,.
\end{equation}
This implies that in the four points $(x_*,y_*) = (\pm 1/\sqrt{2}, \pm 1/\sqrt{2})$ one effectively recovers $w_{\rm eff} \approx 0$, and thus a matter-like evolution of the universe.
Hence the trajectories connecting points $(x_*,y_*)$ with the late time attractor Point~$B$ describe matter to dark energy transitions, and can actually be used as faithful representations of the universe expansion in agreement with the observations.
Note however that a high fine tuning of initial conditions is required in order for this scenario to be physically viable, though in this case the final state of the universe is always an accelerated de Sitter solution.


\subsubsection{$\alpha=0$ and $\beta=1/2$}
Now we analyse a very similar model with a constant scalar field mass, so again $\alpha=0$. But this time we will take $\beta=1/2$, so that the minimum of the scalar field no longer varies in a linear way with the energy density.

\begin{table}
\centering
\begin{tabular}{|c|c|c|c|c|}
  \hline
  Point & $x$ & $y$ & $z$  & $w_{\rm eff}$  \\
  \hline
  \hline
  $A_{\pm}$ & $\pm1$ & 0 & 0 & 1 \\
  \hline
  $B$ & 0 & 0 & $1$ & -1 \\
   \hline
   $C_{\pm}$ & 0 & $\pm 1 $ & 0 & -1 \\
   \hline
   $D$ & 0 & $-\frac{2\gamma}{\sqrt{9+4\gamma^2}}$ & 0 & 0 \\
   \hline
\end{tabular}
\caption{Critical points of the $(0,\frac{1}{2})$-EC dynamical system with a cosmological constant, along with the value of the effective EoS at each point.}
\label{tab:cpL012}
\end{table}

The critical points of the model are displayed in Table~\ref{tab:cpL012}. This time the entire boundary of the $z=0$ plane is not entirely critical, with only the four points $A_{\pm}$ and $C_{\pm}$ lying on this boundary being critical. The system has six critical points:
\begin{itemize}
\item {\it Points $A_{\pm}$. } These points are the standard stiff matter type points dominated by the kinetic energy of the scalar field. The eigenvalues of the system at these points are divergent, however by restricting the dynamics to the $z=0$ plane, and by numerically examining some trajectories in phase space, these points are observed to always be unstable, being either saddle points or the early time attractor.
\item{\it Point $B$.} This point is dominated by the cosmological constant, and represents a de Sitter type expansion with effective EoS $w_{\rm eff}=-1$. The point is the late time attractor, with all trajectories ending up at it.
\item{\it Points $C_{\pm}$.} These points lie on the boundary of the $z=0$ unit circle, and are dominated by the effective potential energy of the scalar field. They are accelerating solutions with an effective EoS $w_{\rm eff}=-1$. The eigenvalues of the Jacobian matrix are divergent at these points, but restricting the dynamics to the $z=0$ plane it is seen that these points are always saddle points, with the Jacobian matrix possessing both positive and negative eigenvalues. 
\item {\it Point $D$. } Point $D$ is a scaling solution, with its effective EoS mirroring that of matter, with $w_{\rm eff}=0$. It exists for all values of the parameter $\gamma$. The Jacobian matrix always has a positive and a negative eigenvalue, and thus the point is always a saddle point. 
\end{itemize}

The system generically evolves from the stiff matter points $A_{\pm}$ to the cosmological constant point $B$, possibly going through the scaling solution $D$ and thus depending on initial conditions, a late time matter to dark energy transition is a viable possibility for this choice of $\alpha$ and $\beta$. 
The $(0,\frac{1}{2})$-EC model, together with a cosmological constant, can thus be used to describe the observed matter to dark energy transition happening in our universe, at least at the background cosmological level.

\subsubsection{$\alpha=1/2$ and $\beta=0$}

Let us now make the parameter choice $\alpha=1/2$ and $\beta=0$. This means that the local minimum of the potential is a constant, however the mass of the scalar field varies proportional to the square root of the energy density. This model is of interest as it has a matter dominated critical point and is simple to analyse mathematically due to it being non-singular everywhere. 

\begin{table}
\centering
\begin{tabular}{|c|c|c|c|c|}
  \hline
  Point & $x$ & $y$ & $z$  & $w_{\rm eff}$  \\
  \hline
  \hline
  $O$ & 0 & 0 & 0 & 0 \\
  \hline
  $A_{\pm}$ & $\pm1$ & 0 & 0 & 1 \\
  \hline
  $B$ & 0 & 0 & $1$ & -1 \\
   \hline
   $C_{\pm}$ & 0 & $\pm 1 $ & 0 & -1/2 \\
   \hline
\end{tabular}
\caption{Critical points of the $(\frac{1}{2},0)$-EC with a cosmological constant along with the value of the effective EoS at each point.}
\label{cpL120}
\end{table}

The critical points are displayed in Table~\ref{cpL120} along with the effective EoS at each of the points. There are six critical points in general. The points $A_{\pm}$ and $B$ behave exactly as in the previous cases: $A_{\pm}$ are either saddle points or the early time attractor, with stiff matter EoS, whereas $B$ is the late time cosmological constant dominated attractor solution. The additional critical points present in this model are:
\begin{itemize}
\item{Point $O$.} This point, the origin of the phase space, is entirely matter dominated with $\sigma=1$ and effective EoS $w_{\rm eff}=0$. The Jacobian matrix at this point always possesses positive and negative eigenvalues, and so is a saddle node for all choices of $\lambda$ and $\gamma$.
\item{Points $C_{\pm}$.} These points are dominated by the effective potential energy of the scalar field. The effective EoS is always accelerating, with $w_{\rm eff}=-1/2$. The eigenvalues of the Jacobian matrix are infinite at these points, but restricting the dynamics to the $z=0$ plane, it is seen that they are always saddle points.
\end{itemize}

\begin{figure}
\centering
\includegraphics[width=0.65\textwidth]{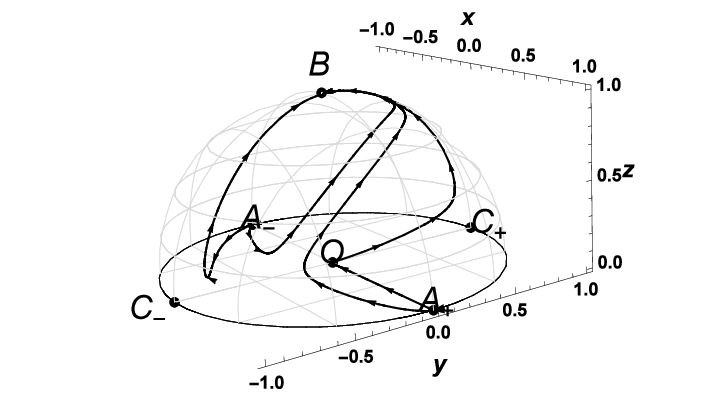}
\caption{Phase space of the dynamical system for the $(\frac{1}{2},0)$-EC with a cosmological constant. The parameter choice $\gamma=1$ and $\lambda=1$ has been made.}
\label{ccalpha12beta0}
\end{figure}
 
In Figure~\ref{ccalpha12beta0} some example trajectories in phase space have been shown for the particular parameter choice $\lambda=1$ and $\gamma=1$. Trajectories start at either of the stiff matter states $A_{\pm}$. They then are drawn towards one or more of the saddle points $C_{\pm}$ and the matter dominated $O$. Eventually all trajectories arrive at the de Sitter expanding late time attractor solution. Thus also this model can exhibit, depending on the initial conditions, a late time dark matter to dark energy transition. 

\subsubsection{$\alpha=1$ and $\beta=-1/2$}

Finally in this section we will consider one more choice for the parameters $\alpha$ and $\beta$. This time we take $\alpha=1$, so that the mass of the scalar field depends linearly on the energy density, and we consider a negative value of $\beta=-1/2$, so that the minimum of the potential depends inversely on the square root of the energy density. In Section~\ref{sub:DS_without_CC}, it was found that even without a cosmological constant late time accelerating attractor solutions were possible, here we extend that analysis to include a cosmological constant. 

\begin{table}
\centering
\begin{tabular}{|c|c|c|c|c|}
  \hline
  Point & $x$ & $y$ & $z$  & $w_{\rm eff}$  \\
  \hline
  \hline
    $O$ & 0 & 0 & 0 & 0 \\
    \hline
  $A_{\pm}$ & $\pm1$ & 0 & 0 & 1 \\
  \hline
  $C_{\pm}$ & 0 & $\pm 1 $ & 0 & 0 \\
  \hline
  $D$ & 0 & $\frac{2\gamma}{\sqrt{9+4\gamma^2}}$ & $\frac{3}{\sqrt{9+4\gamma^2}}$ & -1 \\
   \hline
\end{tabular}
\caption{Critical points of the $(1,-\frac{1}{2})$-EC each dynamical system with a cosmological constant, along with the value of the effective EoS at each point.}
\label{cpL1-12}
\end{table}

The critical points are displayed in Table~\ref{cpL1-12}, along with the effective EoS at each of the points. There are always six critical points: the points $O$, $A_{\pm}$ and $C_{\pm}$ behave exactly the same as in the previous $(\frac{1}{2},0)$-EC model, with the exception that the points $C_{\pm}$ now have effective EoS $w{\rm eff}=0$, and so mirror the effective EoS of the matter dominated solution. The additional new point unique to this model is Point $D$:
\begin{itemize}
\item {\it Point $D$.} This point has an energy contributions from both the interacting potential $f$ and from the cosmological constant. This point exists for all $\gamma$, and in the limit $\gamma\rightarrow0$ it becomes the standard de Sitter type critical point $(0,0,1)$. It too has a de Sitter type effective EoS $w{\rm eff}=-1$. The eigenvalues of the Jacobian matrix evaluated at this point are given by $-3, -3$ and $-3/2$ and so the point is always stable, and is the late time attractor of the system.
\end{itemize}

\begin{figure}
\centering
\includegraphics[width=0.65\textwidth]{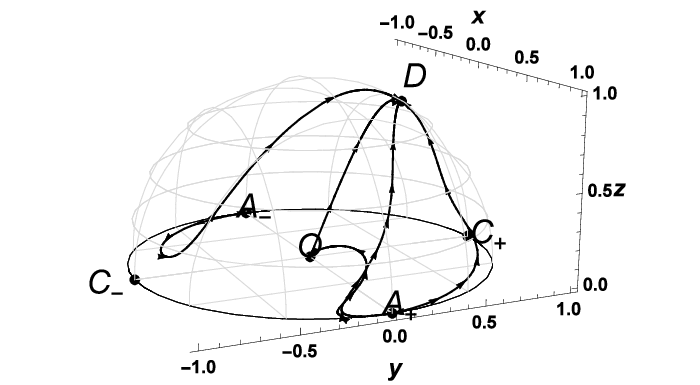}
\caption{Phase space of the dynamical system for the $(1,-\frac{1}{2})$-EC with a cosmological constant. The parameter choice $\gamma=1$ and $\lambda=1$ has been made.}
\label{ccalpha1beta-12}
\end{figure}

Also of interest to note is that the cosmological constant dominated point $(0,0,1)$ is no longer a critical point for this choice of $\alpha$ and $\beta$. The dynamics of the system are very similar to the previous studied case, except now instead of the late time attractor being the cosmological constant solution, it has been replace by the Point $D$, which mirrors the EoS of the de Sitter type solution. In Figure~\ref{ccalpha1beta-12} the phase space of the system is plotted along with some example trajectories. The system starts at either of the stiff matter points $A_{\pm}$, before being drawn towards one of the saddle points $C_{\pm}$ or the matter dominated $O$. All trajectories then arrive at the attractor $D$.

\begin{figure}
\centering
\includegraphics[width=0.55\textwidth]{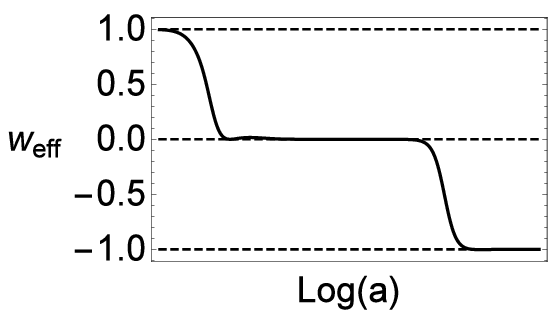}
\caption{Example evolution of $w_{\rm eff}$ for the $(1,-\frac{1}{2})$-EC with a cosmological constant, when $\lambda=1$, $\gamma=1$.}
\label{weffcosmoconstalpha1beta-12}
\end{figure}

The effective EoS of one of the trajectories has been plotted in Figure~\ref{weffcosmoconstalpha1beta-12}; in particular the trajectory which passes through the matter dominated point $O$. The effective EoS begins at the unphysical stiff state with $w_{\rm eff}=1$, before undergoing a period of matter domination, before a dark matter to dark energy transition occurs, ending at a $w_{\rm eff}=-1$. Thus with sufficiently well chosen initial conditions this model can replicate observations. We should also observe that there appears to be no small excursion into the phantom region, as was the case for this model without a cosmological constant (cf.~Fig.~\ref{weffalpha1beta-12}).


\section{Perturbations and structure formation} 
\label{sec:perturbations_and_structure_formation}

In this section we present the cosmological equations of extended chameleons for scalar perturbations at linear order, and discuss their possible effects on structure formation when the quasi-static approximation can be assumed.

\subsection{Perturbation equations}

We choose to work with a standard Newtonian gauge for which the line element can be written as
\begin{align}
  ds^2 = - (1+2\Phi) dt^2+\frac{(1-2\Psi)\, a(t)^2 }{\left[1+\frac{1}{4} K \left(x^2+y^2+z^2\right)\right]^2} \left(dx^2+dy^2+dz^2\right) \,, \label{newtonian}
\end{align}
where $x,y,z$ are Cartesian coordinates and $K = 0, -1, 1$ determines the spatial curvature of the universe.
The scalar perturbations $\Phi$ and $\Psi$ are actually equal, due to the absence of anisotropies in the model under investigation.
In fact the perturbed $ij$-components of the Einstein field equation yield
\begin{align}
  \partial_i\partial_j \left[ \left[1+ \frac{1}{4} K \left(x^2+y^2+z^2\right) \right] (\Psi-\Phi) \right] = 0 \,.
\end{align}
which implies
\begin{align}
  \Phi = \Psi \,.
\end{align}
For this reason in the equations that follow we simply replace $\Phi$ in favour of $\Psi$ whenever it appears.
Before displaying the equations we must first define the perturbed physical quantities: in particular the scalar field, the matter energy density, the matter pressure and the matter four-velocity are perturbed according to
\begin{align}
  \phi + \delta\phi \,, \qquad \rho + \delta\rho \,, \qquad p + \delta p \,, \qquad u_\mu + \delta u_\mu \,, \label{matterpert}
\end{align}
where the last term is defined in terms of the fluid's perturbed scalar velocity $v$ by
\begin{align}
  \delta u_\mu = \left( -\Psi , \partial_i v \right) \,.
\end{align}
At this point we can present the perturbed equations.

The full linear order cosmological equations for Scalar-Fluid theories have been first derived in \cite{Koivisto:2015qua,Boehmer:2015ina}.
In the special case of standard chameleons, where the coupling function $f$ depends only on $n$ through $\rho$ and the scalar field couples only to non-relativistic matter for which $\rho \propto n$ and $p = w \rho = 0$, these equations reduce to 
\begin{multline}
-3 H \dot\Psi +\left[6\frac{K}{a^2}-\frac{k^2 }{a^2}
-\kappa^2 \left(\rho+V+f\right)\right]\Psi \\
  -\frac{\kappa^2}{2}  \left(1+ \frac{\partial f}{\partial \rho}\right) \delta\rho 
-\frac{\kappa^2}{2} \left(\frac{\partial f}{\partial\phi}+V'\right) \delta \phi 
-\frac{\kappa^2}{2}  \dot\phi 
\dot{\delta\phi} = 0 \,, \label{eq:005}
\end{multline}
\begin{gather}
  \dot\Psi +H \Psi + \frac{\kappa^2}{2} \rho \left(1+\frac{\partial f}{\partial \rho}\right) v-\frac{\kappa^2}{2}  \dot\phi \delta
   \phi = 0 \,,
\end{gather}
\begin{multline}
-\frac{\kappa^2}{2} \delta p -\frac{\kappa^2}{2} \rho \frac{\partial^2 f}{\partial \rho^2} \delta\rho
  +\frac{\kappa^2}{2}  \left(\frac{\partial f}{\partial\phi} - \rho \frac{\partial^2 f}{\partial \rho \partial\phi} +V'\right) \delta \phi  \\
   -\frac{\kappa^2}{2}  \dot\phi \dot{\delta\phi} +\left( 2\dot{H} +3H^2+\frac{\kappa^2}{2}\dot\phi^2 -\frac{K}{a^2} \right) \Psi
  +4 H \dot\Psi+\ddot\Psi = 0 \,, \label{eq:006}
\end{multline}
\begin{gather}
  \ddot{\delta\phi}+3 H \dot{\delta\phi}+\left(\frac{k^2}{a^2}+\frac{\partial^2 f}{\partial\phi^2} 
+V''\right) \delta \phi +\frac{\partial^2 f}{\partial \rho \partial\phi} \delta\rho 
   -2 \left(\ddot\phi+3H\dot\phi\right) \Psi -4 \dot\phi \dot\Psi = 0 \,, \label{eq:008}
\end{gather}
\begin{gather}
  \dot{\delta\rho} +3H\left(\delta\rho + \delta p \right) -\rho \left(\frac{k^2}{a^2} v +3 \dot\Psi \right) =0 \,,  \label{eq:007}
\end{gather}
\begin{multline}
\left(1+ \frac{\partial f}{\partial \rho}\right)\dot{v}
  +\left( \dot\phi \frac{\partial^2 f}{\partial \rho \partial\phi} -3 H \rho 
\frac{\partial^2 f}{\partial \rho^2}\right) v 
  +\frac{\partial^2 f}{\partial \rho^2} \delta\rho+\frac{\partial^2 
f}{\partial \rho \partial\phi} \delta \phi
   +\frac{1}{\rho}  \delta p
  +\left(1+ \frac{\partial f}{\partial \rho}\right) \Psi = 0 \,, \label{eq:004}
\end{multline}
where we passed to Fourier space substituting the spatial Laplacian with the wave number of the fluctuation: $\nabla^2 \mapsto -k^2$.
Of course in Eqs.~\eqref{eq:005}--\eqref{eq:004} one should consider that the coupling function $f(\rho,\phi)$ of extended chameleons is given by \eqref{eq:ext_cham_coupl}, namely
\begin{equation}
  f(\rho,\phi) = \frac{1}{2} m^2(\rho) \left[ \phi - \phi_0(\rho) \right]^2 \,,
\end{equation}
with the functions $m^2(\rho)$ and $\phi_0(\rho)$ defined as
\begin{equation}
  m^2(\rho) = A\, \rho^\alpha  \quad\mbox{and}\quad \phi_0(\rho) = B\, \rho^\beta \,.
\end{equation}

\subsection{Growth of structures in the quasi-static approximation}

Equations~\eqref{eq:005}--\eqref{eq:004} can be used to study the implications of extended chameleons in several cosmological contexts.
We are particularly interested in the structure formation process, where the quasi-static limit can be applied and the evolution of (non-relativistic) matter overdensities $\delta$, in the comoving matter gauge, is uniquely determined by the following equations (see \cite{Koivisto:2015qua} for details)
\begin{eqnarray} \label{eq:ddhat}
\ddot{\delta}  +  \lb \lp 2 + 3c_s^2 \rp H + 2Y\dot{\phi} \rb \dot{\delta} 
+ c_s^2 \lp \frac{k}{a} \rp^2 {\delta} 
 =   \lp C_0  + C_1 H\dot{\phi} + C_2 \dot{\phi}^2\rp{\delta}\,,
\end{eqnarray}
where
\begin{eqnarray}
C_0 & = & \frac{\kappa^2}{2} \lp  \rho +  4c_s^2 
\dot{\phi}^2\rp - 9c_s^2 H^2 + \frac{2}{3} \frac{K}{a^2} 
+  \lb X^2 - \lp 1+c_s^2\rp XY + Y^2 \rb \rho - \lp X-Y\rp V' \,, \qquad \\
C_1 & = &  \lp 5 + 6 c_s^2\rp X - \lp 5 +3c_s^2\rp Y -3 R \,, \\
C_2 & = & U^2 - X^2 - XY -4 Y^2\,,
\end{eqnarray}   
and where we have defined the quantities
\begin{equation} \label{qder}
      X \equiv \frac{\rho}{\rho+f} \frac{\partial^2 f}{\partial \rho \partial \phi} \,,
\quad Y \equiv \frac{1}{\rho+f} \frac{\partial f}{\partial \phi} \,, 
\quad U \equiv \left( \frac{\partial^3 f}{\partial\phi^2 \partial \rho} \right)^{1/2} \,, 
\quad R \equiv \rho \frac{\partial^3 f}{\partial\rho^2 \partial \phi} \,,
\end{equation}
together with the sound speed square of the fluid in the rest frame of the field
\begin{equation} \label{cs2}
c_s^2 \equiv \rho \frac{\partial^2 f}{\partial\rho^2} \left({1+\frac{\partial f}{\partial \rho}}\right)^{-1} \,.
\end{equation} 

All the expressions \eqref{qder}--\eqref{cs2} are solely defined in terms of background quantities, and can thus be rewritten in terms of the dimensionless variables \eqref{dynamsys:variables} and \eqref{eq:z_var_CC} used in Sec.~\ref{sub:DS_with_CC} for the dynamical systems analysis with a cosmological constant.
In what follows we will only focus on EC models where a non-vanishing cosmological constant appears. As we found in Sec.~\ref{sec:dynamical_systems_analysis}, these are in fact the scenarios where the observed matter to dark energy transition might be reproduced in almost all situations.
More quantitatively for the quantities \eqref{qder} we find
\begin{align}
  X &= \frac{1}{ y^2 + \sigma^2} \left[ \alpha \sqrt{2A} y \sigma^\alpha \left(\frac{z}{\sqrt{\Lambda}}\right)^{1- \alpha} - \beta AB \sigma^{2(\alpha + \beta)} \left(\frac{z}{\sqrt{\Lambda}}\right)^{2 (1- \alpha - \beta)} \right] \,, \label{eq:X} \\
  Y &= \sqrt{2A} \frac{y \sigma^\alpha}{y^2 + \sigma^2} \left(\frac{z}{\sqrt{\Lambda}}\right)^{1- \alpha} \,,\\
  U &= \sqrt{\alpha A} \frac{\sigma^\alpha}{\sqrt{y^2 + \sigma^2}} \left(\frac{z}{\sqrt{\Lambda}}\right)^{1- \alpha} \,,\\
  R &= \frac{1}{y^2+ \sigma^2} \left[ \alpha (\alpha-1) \sqrt{2A} y \sigma^\alpha \left(\frac{z}{\sqrt{\Lambda}}\right)^{1- \alpha} - (2 \alpha + \beta -1) \beta AB \sigma^{2(\alpha + \beta)} \left(\frac{z}{\sqrt{\Lambda}}\right)^{2(1- \alpha -\beta)} \right] \,, \label{eq:R}
\end{align}
while for the sound speed \eqref{cs2} we have
\begin{multline}
  c_s^2 = \Bigg[ \alpha (\alpha-1) y^2 -2 \alpha \beta \sqrt{2A}B \sigma^{\alpha + 2 \beta} \left(\frac{z}{\sqrt{\Lambda}}\right)^{1- \alpha -2 \beta} + \beta^2 AB^2 \sigma^{2(\alpha +2 \beta)} \left(\frac{z}{\sqrt{\Lambda}}\right)^{2(1- \alpha -2 \beta)} \\ 
  -\beta (\beta-1) \sqrt{2A} B y \sigma^{\alpha +2 \beta} \left(\frac{z}{\sqrt{\Lambda}}\right)^{1- \alpha -2 \beta} \Bigg] / \Bigg[ \sigma^2 + \alpha y^2 - \beta \sqrt{2A} B y \sigma^{\alpha +2 \beta} \left(\frac{z}{\sqrt{\Lambda}}\right)^{1- \alpha -2 \beta} \Bigg] \,.
  \label{eq:cs2_cham}
\end{multline}

We can now check how Eq.~\eqref{eq:ddhat} reduces in the effectively matter dominated critical points obtained in Sec.~\ref{sub:DS_with_CC} for different extended chameleon models, particularly the ones where a matter to dark energy transition is indeed possible.
These matter-like evolving points should in fact describes the epoch of the universe during which structures form.
Any deviation in Eq.~\eqref{eq:ddhat} from $\Lambda$CDM, for which the quantities \eqref{eq:X}--\eqref{eq:cs2_cham} are all zero, could thus in principle provide physical effects that might lead to observable signatures to look for in different cosmological datasets.
In what follows we assume a spatially flat universe in agreement with the observations, and thus we set $K = 0$. 

\subsubsection{$\alpha = 0$ and $\beta = 1$} 

We first analyse the extended chameleon model with $\alpha = 0$ and $\beta = 1$, namely $(0,1)$-EC.
The physical implications of this model at Solar System and galactic scales have been investigated in \cite{Brax:2015fcf}, where it was shown that interesting effects in the galaxy rotation curves can arise while satisfying all smaller scales constraints.
Moreover this model represents a new kind of screening where the mass of the scalar field remains constant, while the only function that varies when the energy density of matter changes is the minimum of the effective potential, namely $\phi_0$.

According to the background analysis performed in Sec.~\ref{sub:DS_with_CC}, the phase space of $(0,1)$-ECs does not explicitly present any matter dominated critical points.
However, as noted above, the points $(x,y,z) = (\pm 1/\sqrt{2},\pm 1/\sqrt{2},0)$ effectively constitute matter dominated solutions since, assuming $\sigma^2 \gg z$, they yield $w_{\rm eff} \approx 0$.
In order to understand if such points can indeed characterize the observed matter dominated era of our universe, at least at sub-horizon scales, we must check if deviations are present in Eq.~\eqref{eq:ddhat}.
Setting $\alpha=0$ and $\beta=1$ in the quantities \eqref{eq:X}--\eqref{eq:cs2_cham} we find
\begin{equation}
  X = -\frac{A B \sigma ^2}{\sigma ^2+y^2} \,, \quad
  Y = \frac{\sqrt{2} \sqrt{A} y z}{\sqrt{\Lambda } \left(\sigma ^2+y^2\right)} \, \quad
  U = 0 \,, \quad
  R = 0 \,, \quad
  c_s^2 = \frac{A B^2 \Lambda  \sigma ^2}{z \left(z- \sqrt{2} \sqrt{A} B \sqrt{\Lambda } y\right)} \,.
\end{equation}
Recalling the assumption $\sigma^2 \gg z$, in the points $(x,y,z) = (\pm 1/\sqrt{2},\pm 1/\sqrt{2},0)$ these expressions all vanish.
This implies that for $(0,1)$-ECs there are no signs of deviations from $\Lambda$CDM in the growth of structure.
The matter to dark energy transition from any of the points $(\pm 1/\sqrt{2},\pm 1/\sqrt{2},0)$ to the cosmological constant dominated late time attractor $(0,0,1)$, represents thus a faithful description of the observed universe and it is indistinguishable from the $\Lambda$CDM dynamics, at least within the approximations that we have assumed.
Note that this result is obtained regardless of the fact that in points $(\pm 1/\sqrt{2},\pm 1/\sqrt{2},0)$ the universe is actually dominated by the scalar field and not by the (dark) matter component.


\subsubsection{$\alpha = 0$ and $\beta = 1/2$} 

The second model we consider is defined by $\alpha = 0$ and $\beta = 1/2$, and it is denoted by $(0,\frac{1}{2})$-EC.
This model is similar to $(0,1)$-EC since one still obtains a constant scalar field mass with an environment dependent minimum, effectively belonging to a new class of screening mechanisms.
Unfortunately it is not as simple as $(0,1)$-EC to find analytical solutions in a galactic framework characterized by an NFW profile, and thus the implications on galaxy rotations curves for $(0,\frac{1}{2})$-EC have not been investigated in \cite{Brax:2015fcf}.

For $\alpha = 0$ and $\beta = 1/2$ we found in Sec.~\ref{sub:DS_with_CC} that there is a saddle critical point at $(x,y,z) = (0, - 2 \gamma / \sqrt{9+4 \gamma^2}, 0)$; namely Point~$D$ in Table~\ref{tab:cpL012}.
This point represents a matter scaling solution since the energy density of the scalar field does not vanish, though its kinetic energy is zero, and the effective EoS of the universe mimics a matter expansion: $w_{\rm eff} = 0$.
Moreover in this model a dark matter to dark energy transition is always possible since the late-time attractor of the phase space is always a de Sitter solution, effectively reproducing a cosmological constant driven accelerated expansion.

We want again to analyse deviations from $\Lambda$CDM in Eq.~\eqref{eq:ddhat}, assuming the universe is well described by point~$D$ during its matter dominated epoch.
Using the values $\alpha=0$ and $\beta=1/2$ the quantities \eqref{eq:X}--\eqref{eq:R} become
\begin{equation}
  X = -\frac{AB}{2\sqrt{\Lambda}} \frac{z \sigma}{y^2+\sigma^2} \,, \quad
  Y = \sqrt{\frac{2A}{\Lambda}} \frac{y z}{y^2 + \sigma^2} \,, \quad
  U = 0 \,, \quad
  R = \frac{A B}{4 \sqrt{\Lambda}} \frac{z \sigma}{y^2 + \sigma^2} \,,
  \label{eq:001}
\end{equation}
while the sound speed \eqref{eq:cs2_cham} reduces to
\begin{equation}
  c_s^2 = \frac{1}{4} \frac{ A B^2 \sigma+\sqrt{2 A} B  y}{\sigma-\frac{\sqrt{A} B  y}{\sqrt{2}}} \,.
\end{equation}
In the scaling solution $D=(0, - 2 \gamma / \sqrt{9+4 \gamma^2}, 0)$ all expressions in \eqref{eq:001} vanish, while the sound speed becomes
\begin{equation}
  c_s^2 = \frac{\sqrt{2A} B}{4} \left( \frac{3 \sqrt{2A} B - 2 \gamma }{3 + \sqrt{2A} B \gamma }  \right) = 0 \,,
\end{equation}
where in the last equality we substituted $\gamma$ in terms of $B$, as given by Eq.~\eqref{eq:params_cc}.
We thus find that in this case there are no deviations from $\Lambda$CDM in Eq.~\eqref{eq:ddhat}, implying that no observational signatures will be present at sub-horizon scales in the $(0,\frac{1}{2})$-EC model.




\subsubsection{$\alpha = 1/2$ and $\beta = 0$} 

The third model we deal with is $(\frac{1}{2},0)$-EC.
As shown in \cite{Brax:2015fcf}, the fifth force between two microscopic test masses and all the effects within the Solar System, are identically zero for this model.
This implies that no constraints can be forced upon it from small scales experiments.
Nevertheless the dynamics at large scale might present deviations from $\Lambda$CDM and possible observational effects.

In Sec.~\ref{sub:DS_with_CC} we have found that the standard matter dominated point $(0,0,0)$ is a saddle critical point for $(\frac{1}{2},0)$-EC.
Together with the fact that the late time attractor can be given by a dark energy dominated critical point (either Point~$B$ or Points~$C_\pm$ in Tab.~\ref{cpL120}), also in this model the background cosmological dynamics can match the one reconstructed by observations.
Again we want to see if this is also the case at the perturbation level by checking Eq.~\eqref{eq:ddhat}.
Setting $\alpha=1/2$ and $\beta=0$ in Eqs.~\eqref{eq:X}--\eqref{eq:cs2_cham} we obtain
\begin{gather}
  X = Y = \sqrt{\frac{A}{2\sqrt{\Lambda}}} \left(\frac{ y  \sqrt{z \sigma}}{\sigma ^2+y^2}\right) \,, \quad
  U = \sqrt{\frac{A}{2\sqrt{\Lambda}}} \left(\frac{ \sqrt{z \sigma}}{\sigma ^2+y^2}\right) \,, \nonumber \\
  R = - \frac{1}{2} \sqrt{\frac{A}{2\sqrt{\Lambda}}} \left(\frac{ y  \sqrt{z \sigma}}{\sigma ^2+y^2}\right) \,, \quad
  c_s^2 = -\frac{1}{2} \left( \frac{y^2}{2 \sigma ^2+ y^2} \right) \,.
\end{gather}
At the matter dominated point $(0,0,0)$ all these quantities vanish and thus no deviations from $\Lambda$CDM are present in $(\frac{1}{2},0)$-EC, neither at the background nor at the perturbations level, at least at sub-horizon scales.
Extended chameleons models for which $\beta = 0$ are in fact able to effectively hide the presence of the scalar field at all scales.
As already noticed in \cite{Brax:2015fcf}, this is related to the fact that any screening mechanism for which the minimum of the effective potential does not depend on the matter energy density, cannot give rise to a physically meaningful fifth force.


\subsubsection{$\alpha = 1$ and $\beta =-1/2$} 

The last example we consider is $(1,-\frac{1}{2})$-EC, where we allow for a negative value of beta, describing a minimum $\phi_0$ inversely proportional to (the square root of) the matter energy density $\rho$.
In Sec.~\ref{sub:DS_with_CC} we found that this model not only present a standard matter dominated critical point at $(0,0,0)$, but its phase space contains also two scalar field solutions at the points $(0,\pm 1,0)$ which mimic a matter dominated expansion.
A matter to dark energy transition in agreement with observations can be obtained from a trajectory connecting any of these points to the dark energy dominated late time attractor.
For this reason we now look at how Eq.~\eqref{eq:ddhat} behaves in all these three matter critical points.

First of all we need to see how the quantities \eqref{eq:X}--\eqref{eq:cs2_cham} reduce for $(1,-\frac{1}{2})$-EC.
Setting $\alpha = 1$ and $\beta = -1/2$ we obtain
\begin{equation}
  X = \frac{\frac{A B}{2 \sqrt{\Lambda }} \sigma z +\sqrt{2A} \sigma  y}{\sigma ^2+y^2} \,, \quad
  Y = \sqrt{2A} \left( \frac{\sigma  y}{\sigma ^2+y^2} \right) \,, \quad
  U = \frac{\sqrt{A} \sigma }{\sqrt{\sigma ^2+y^2}} \,, \quad
  R = \frac{A B \sigma  z}{4 \sqrt{\Lambda } \left(\sigma ^2+y^2\right)} \,,
\end{equation}
while the sound speed becomes
\begin{equation}
  c_s^2 = \frac{\sqrt{A} B z \left[\sqrt{A} B z+\sqrt{2 \Lambda} (4-3 y)\right]}{2 \sqrt{2A \Lambda}  B  y z+4 \Lambda  \left(\sigma^2+y^2\right)} \,.
\end{equation}
In the scalar field solutions $(0,\pm 1,0)$ one can easily check that all these expressions are zero.
This implies that for these solutions the small scale dynamics of structure formation is equivalent to the $\Lambda$CDM one, irrespectively of the fact that the cosmic evolution is completely driven by the scalar field.
On the other hand, in the matter dominated point $(0,0,0)$ all these expressions vanish, but $U$ which simply becomes a constant: $U = \sqrt{A}$.
In this situation Eq.~\eqref{eq:ddhat} becomes
\begin{equation}
  \ddot\delta + 2 H \dot\delta = \left( \frac{\kappa^2}{2} \rho + A \dot\phi^2 \right) \delta \,,
\end{equation}
which, dividing by $H^2$, can be rewritten as
\begin{equation}
  \ddot\delta + 2 H \dot\delta = 3 H^2 \left( \frac{1}{2} \sigma^2 + \frac{2 A}{\kappa^2} x^2 \right) \delta \,.
  \label{eq:009}
\end{equation}
From this last equation one can realize that no effects of the scalar field will appear at sub-horizon scales if the background evolution is assumed to be described by point $(0,0,0)$ where $x=0$.
If this approximation is relaxed, i.e.~if one considers a more realistic trajectory in the phase space passing near the origin where $x$ is small but not zero, then the last term in Eq.~\eqref{eq:009} could actually provide some interesting deviations from the standard $\Lambda$CDM result, possibly leading to interesting observational signatures.
This would however require a full numerical investigation at both background and perturbation levels, which lies outside the scope of the present analysis and will be thus left for future works.



\section{Conclusions} 
\label{sec:conclusions}

In this work we have analysed the cosmological dynamics of extended chameleons (ECs), at both background and linear perturbation levels.
This class of scalar field theories has been found to be particularly useful not only to define new and generalise old screening mechanisms at Solar System scales, but its phenomenological implications at galactic distances can also provide interesting observational signatures to compare against astronomical data \cite{Brax:2015fcf}.

To investigate the cosmological background evolution we have employed dynamical systems techniques, which allowed us to obtain the whole phase space of specific EC models and thus to characterize their dynamical properties completely.
We have first considered EC models with a vanishing cosmological constant in order to understand if the scalar field is able to drive the late time acceleration of the universe by itself (Sec.~\ref{sub:DS_without_CC}).
Unfortunately this is not the case in general, and only for some specific models one can find an early time matter dominated solution followed by a late time dark energy dominated attractor.
For this reason we have also studied the cosmological dynamics of EC models when a positive cosmological constant is present (Sec.~\ref{sub:DS_with_CC}).
In this scenario, the phase space of any EC model contains a late time attractor where the universe is dominated by the cosmological constant and it undergoes a de Sitter accelerated expansion.
Moreover for almost all of the EC models there also exists an effectively matter dominated saddle point, which can well represent the early time epoch of the universe where large scale structures form.
Depending on the specific EC model, these matter points can be either characterized by the (dark) matter component or by the scalar field, with some of them even representing scaling solutions, where the energy densities of both matter and dark energy are of the same order of magnitude and scale equally with time.
Any trajectory connecting any of these matter points to the accelerated future attractor, well represents the observed late time evolution of the universe, at least at the background level (cf.~e.g.~Fig.~\ref{weffcosmoconstalpha1beta-12}).

In order to check the consistency of these solutions beyond the background framework, we have then analysed the dynamics of linear cosmological perturbations at sub-horizon scales, adopting in particular the quasi-static approximation (Sec.~\ref{sec:perturbations_and_structure_formation}).
As a working assumption we have supposed that the matter dominated epoch of the universe is well described at the background level by any of the effectively matter dominated critical points derived in the previous section, ignoring in this way the transition to dark energy happening at late times.
Within these approximations we have found that no EC models, among the ones considered here, imply any physical deviation from the standard result obtained within $\Lambda$CDM, and thus the sub-horizon formation of structures can take place in agreement with the observations.
Nevertheless it might well be that other applications of the linear cosmological perturbations equations of ECs presented in this work, provide possible observational signatures that will allows us to distinguish the dynamics of these scalar field models from the one of $\Lambda$CDM, and eventually to constrain them with present and future probes.
For example, a full numerical analysis of EC models at the perturbation level, taking into account also the evolution from matter to dark energy domination, can actually give small inconsistencies with respect to $\Lambda$CDM, especially at late times.
Furthermore an investigation of the perturbations at the non-linear level, for example assuming spherical over-densities, could highlight some effects that are not present at the linear level and fully exploits the properties of the EC models at the cluster scales, in analogy with the studies performed e.g.~for $K$-mouflage \cite{Brax:2014yla,Brax:2014gra,Barreira:2014gwa,Brax:2015lra} and symmetron \cite{Davis:2011pj,Brax:2011pk,Brax:2012nk} theories.
Such analyses reside however beyond the scopes of the present paper and are thus left as material for future works.

To conclude we have shown that EC models can well represent alternative descriptions of the observed evolution of the universe at late times.
In fact, concerning the applications analysed in this work, the cosmological dynamics of these scalar field models is indistinguishable from the one of $\Lambda$CDM, at both background and perturbation levels.
This result implies that EC theories can consistently be employed as models of dark energy in agreement with the observations at both large, i.e.~cosmological, and small, i.e.~Solar System, distances.
As shown in \cite{Brax:2015fcf} they can however provide some astrophysical implications at intermediate, i.e.~galactic, scales, where interesting signatures can arise in the galaxy rotation curves, and thus be compared with observational data.
It will be the work of future investigations to derive physical constraints on EC models and to further characterizes their phenomenology at all scales.

\subsection*{Acknowledgements} 
\label{sub:acknowledgements}

NT acknowledge support from the Labex P2IO and the Enhanced Eurotalents Programme.



\bibliographystyle{unsrt}
\bibliography{bibfile}

\end{document}